\documentclass[11pt]{article}

\usepackage{float}
\usepackage{cite}
\usepackage{siunitx}
\usepackage{orcidlink}
\usepackage{lmodern}
\usepackage{indentfirst}
\usepackage{authblk}
\usepackage{xurl}
\usepackage{hyperref}
\DeclareSIUnit{\angstrom}{\text{\AA}}

\newcommand{\BHmax}{BH_{max}}

\title{Tailoring the material properties, nanostructure and grain alignment of Alnico magnets through micromagnetic simulations.}

\author[1,5]{Anda Elena Stanciu\,\orcidlink{0000-0002-8717-5596}\thanks{Corresponding author \newline Email-address: anda.stanciu@donau-uni.ac.at \newline Preprint submitted to IOPscience}}
\author[1]{Johann Fischbacher\,\orcidlink{0000-0002-0815-5379}}
\author[1]{Markus Gusenbauer\,\orcidlink{0000-0002-3540-3964}}
\author[1]{Alexander Kovacs\,\orcidlink{0000-0002-8413-1950}}
\author[1]{Harald Oezelt\,\orcidlink{0000-0002-3754-3565}}
\author[2]{Joachim Seland Graff\,\orcidlink{0000-0001-6072-1334}}
\author[2]{Patricia Carvalho\,\orcidlink{0000-0002-5447-0409}}
\author[3]{Anette Eleonora Gunnæs\,\orcidlink{0000-0002-2675-0411}}
\author[4]{Matej Zaplotnik\,\orcidlink{0000-0000-0000-0000}}
\author[2]{Espen Sagvolden\,\orcidlink{0000-0002-2268-1260}}
\author[2]{Spyros Diplas\,\orcidlink{0000-0002-9295-464X}}
\author[1]{Thomas Schrefl\,\orcidlink{0000-0002-0871-0520}}

\affil[1]{Department for Integrated Sensor Systems, University for Continuing Education Krems, Wiener Neustadt 2700, Austria}
\affil[2]{Materials Physics Oslo, Sustainable Energy Technology, SINTEF Industry, Forskningsveien 1, 0373 Oslo, Norway}
\affil[3]{Department of Physics, University of Oslo (UiO), 0316 Oslo, Norway}
\affil[4]{Magneti Ljubljana, d.d., 1000 Ljubljana, Slovenia}
\affil[5]{National Institute of Materials Physics, Atomistilor 405A, RO 077125 Magurele, Romania}

\date{}

\begin{document}

\maketitle

\affil{$^*$Anda Elena Stanciu}



\begin{abstract}
Alnico magnets have gained renewed interest in the search for rare-earth free permanent magnets due to their high thermal stability and magnetisation. However, the limited coercivity of these shape-anisotropy-based alloys constrains their performance. Their coercivity can be improved by tailoring their structure at the micro- and nano-levels and by identifying the elemental composition and crystal structure that would increase anisotropy. Alnico alloys are composed of millimeter-to-micrometer-sized grains of different sizes and orientations. Inside each grain, magnetic nanorods embedded in a non-magnetic matrix form as a result of thermo-magnetic treatment of the cast alloy. Starting from a reference Alnico sample, we realised a finite elements micromagnetic study of exchange-decoupled rods by varying their dimensions and interrod spacing across those observed experimentally. We computed the hysteresis properties by progressing from micromagnetic simulations of a small number of rods within the magnetostatic field of their neighbours to large systems treated statistically based on the distribution of orientations of the grains. We compared the coercivity of an isolated rod with that of the exchange-decoupled system to highlight the effect of magnetostatic interactions. We computed analytically the stray field acting on a single rod as a consequence of its surrounding rods in order to confirm the scaling of the coercivity with the packing fraction $p$. We explored how intrinsic material properties influence magnetic behaviour by examining materials with different magnetocrystalline anisotropy constants and saturation polarisation values. Results from several hundred simulations were used to train a multi-layer perceptron regressor and predict the magnetic properties as function of the dimensions of the rods, interrod spacing and orientation of the grains. With this approach, we highlight the underlying trends by which nanoscale structuring, intrinsic material properties and grain alignment can be tailored to improve the magnetic properties of Alnico alloys. 

\end{abstract}

\section{Introduction}
In the context of the growing global demand for emerging environmentally friendly energy and electric mobility applications \cite{Lewis_RevPermMag2013, Mohapatra_HandbookMagMat2018, Poudel_IEEEMagnetics2021, Eklund_IEEE2020}, there has been renewed interest in improving the performance of rare-earth elements (REE) free permanent magnets \cite{Kovacs_Engineering2020} based also on novel preparation methods available, e.g., additive manufacturing. Alnico alloys are notable for being REE-free and for possessing high Curie temperature and remanent magnetisation \cite{Zhou_ActaMaterialia2014, Zhang_JMMM2018}. Their good thermal stability makes them suitable for high-temperature applications.   

For a material to function as a promising permanent magnet, it must possess good intrinsic properties (spontaneous magnetisation and uniaxial magnetocrystalline anisotropy) and a tailored microstructure designed to resist magnetisation reversal through controlled nucleation and pinning mechanisms \cite{CoeySkomsky1999, Fischbacher_JPhysD2018}. At microscopic level, representative Alnico alloys consist of an ensemble of grains with different sizes and orientations where the alignment of the magnetic rods within a grain is much more pronounced than between grains \cite{Zhou_ActaMaterialia2014}. The nanostructure of Alnico alloys that underpins their magnetic properties is obtained by thermomagnetic treatment of a cast alloy of Fe, Co, Ni, and Al with minor additions of Cu and Ti \cite{McCurrie_1982}. During cooling, the alloy undergoes the spinodal decomposition that would ideally result into Fe-Co-rich magnetic nanorods (phase $\alpha_1$) embedded in a non-magnetic matrix (phase $\alpha_2$) \cite{Zhou_ActaMaterialia2018}. 

The coercivity $H_c$ is determined by anisotropies present. For magnetic particles of small sizes, with uniform magnetisation that rotates coherently under an applied field, the Stoner–Wohlfarth model applies, where the energy of the system is given by the uniaxial anisotropy and by the Zeeman energy \cite{Luborsky_JAppPhys1957, Chikazumi_PhysMag1964}. The uniaxial anisotropy may originate from both magnetocrystalline and shape anisotropy contributions: $H_c = \frac{2K_1}{\mu_0 M_s} + \frac{M_s}{2}(1-3N)$ , where $K_1$ is the magnetocrystalline anisotropy constant, $\mu_0$ is the vacuum permeability, $M_s$ is the saturation magnetisation and $N$ is the demagnetisation factor parallel to the $c$-axis \cite{Coey_2010}. In Alnico alloys, the low magnetocrystalline anisotropy makes coercivity dependent mainly on the shape anisotropy \cite{Bance_JMMM2014, Ke_APL2017, Kronmuller_2003} that can be tuned by the preparation conditions. For instance, the length of the magnetic nanorods can be directly enhanced by increasing the intensity of the external magnetic field applied during the thermo-magnetic treatment \cite{McCurrie_1982, Zhou_ActaMaterialia2014, Zhao_JMMM2023}. Despite the theoretical limit of 1.05 T of the coercivity of Alnico magnets, the demagnetising fields, thermal fluctuations, and misorientation of the magnetic nanorods with respect to the alignment direction of the magnet reduce the coercivity determined for ideal structures \cite{Mohapatra_PhysRevB2022, Fischbacher_ApplPhysLett2017}, the highest value obtained experimentally being around $\SI{0.2} {T}$ \cite{White_AppSci2019}. The Alnico magnets are typically characterised by their maximum energy density product ($\BHmax$), which is calculated as the area of the largest rectangle that can be accommodated under the $B-H$ curve. A good permanent magnet material requires a high spontaneous magnetisation and a coercivity higher than half of the spontaneous magnetisation. If these conditions are satisfied, $\BHmax \propto M_s^2$. If coercivity is less than half of the remanent magnetisation, the maximum energy density product is limited by coercivity \cite{Fischbacher_JPhysD2018}. We note that the highest maximum energy density product achieved experimentally for Alnico magnets is of approx. $\SI{80}{kJ/m^3}$~\cite{Anderson_AIPAdv2017}. We will therefore discuss coercivity, remanent magnetisation, and maximum energy density product extracted from the simulation of a reversal process as the investigated hysteresis properties.

The design of Alnico magnets has been extensively investigated by computations \cite{FidlerJPhysD_2000, Ke_APL2017, Won_IEEE2021, Schabes_JMMM1991} and experiments \cite{Zhou_ActaMaterialia2017, Zhou_JALCOM2025, Rehman_JALCOM2025, Rahimi_JMatEngPerf2025, Dussa_AppMatToday2025} in the search for optimal compositional and morpho-structural properties. However, recent advances in computational modelling and synthesis techniques support efforts to improve the coercivity of Alnico magnets \cite{Bonthu_IEEE2017, Cui_ActaMaterialia2018}. Our investigation utilised a reference Alnico5-7 sample  with $\BHmax = \SI{60}{kJ/m^3}$ as a baseline for comparison with the micromagnetic simulation results. In the following, we will call it reference sample. We characterised the morphostructural and magnetic properties of this sample and subsequently realised a micromagnetic modelling study of nanostructured magnetic rods with intrinsic material properties, dimensions, spatial arrangement of rods in a grain and the disorder in-between grains ranging around those of the reference sample. 

\begin{figure}[H]
 \centering
        \includegraphics[width=0.98\textwidth]{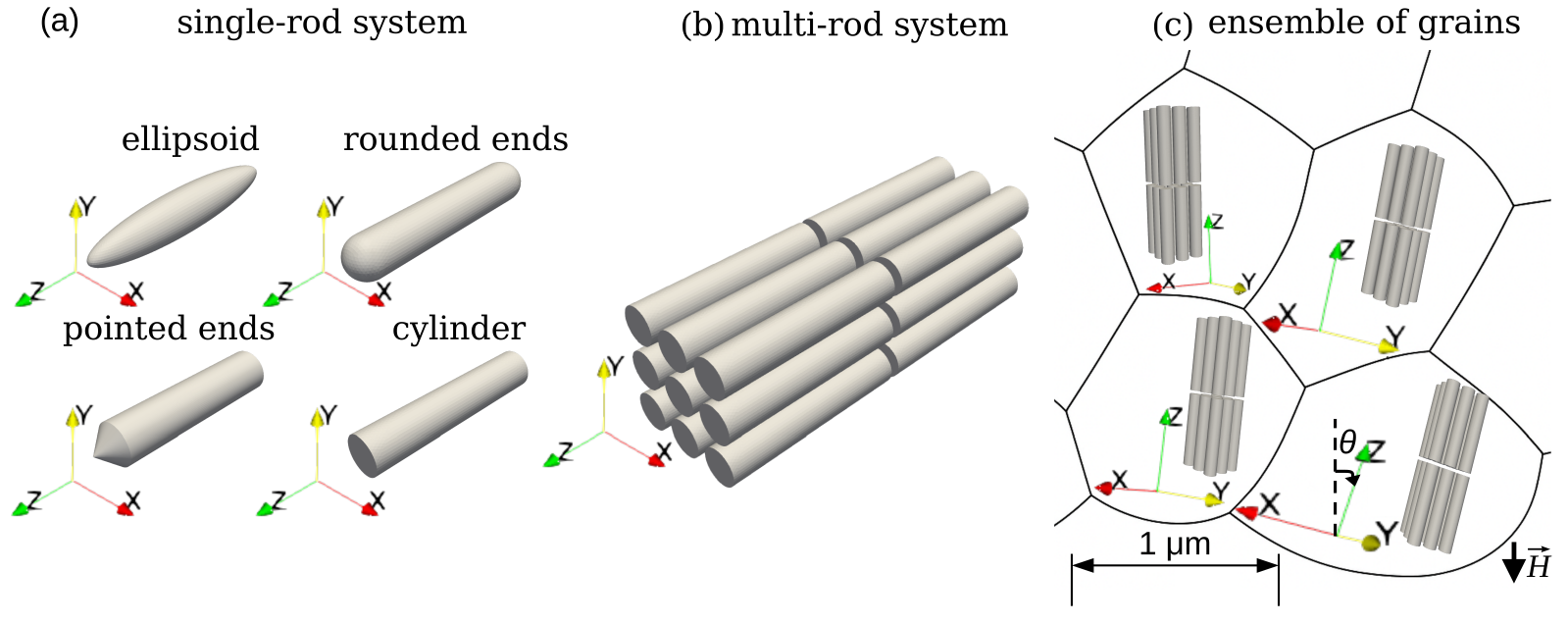}
 \caption{The single-rod system shaped either as an ellipsoid or as a cylinder with flat, rounded or pointed ends (a). The multi-rod system consisting of cylindrical rods (b). Schematic representation of millimeter-to-micrometer-sized grains showing the disorder between grains characterised by the misorientation angle $\theta$ between the $z$-axis of the rods and the applied field (c).} 
\label{fig1}
\end{figure}

We first investigated the magnetic behaviour of an isolated rod with different shapes which will be further referred to as the single-rod system. We modelled a single grain as a multi-rod system consisting of 18 rods distributed across two layers, with each layer comprising nine rods. The number of rods in x- and y-directions is 3, whereas the number of rods in the z-direction is 2. We present a schematic illustration of the setups used to simulate the single and multi-rod systems in Fig. \ref{fig1} (a) and (b), respectively. The long axes of the rods are aligned with the $z$-axis. Fig. \ref{fig1} (c) shows the orientation of the rods within individual grains with respect to the applied field. The simulation results were fed into a machine learning model to predict hysteresis properties as a function of rod diameter, aspect ratio (defined as rod length divided by its diameter), interrod spacing, misalignment and intrinsic material properties. We validated our simulation results against experimental data and propose pathways for optimising the nanostructure and material properties to enhance the performance of the studied Alnico permanent magnet.
We describe the experimental procedures and the micromagnetic modelling framework in Section 2. We present the results of microstructural characterisation of the reference sample, the micromagnetic modelling of the single and multi-rod systems and the statistical treatment of coercivity of the ensemble of grains in Section 3, as described in the following. We utilised the microstructural properties of the reference sample to define the relevant rod dimensions, aspect ratios and grain morphology. We analysed first the isolated rods and assessed the influence of rod shape on the magnetisation reversal and coercivity. Next, we investigated magnetostatic interactions between rods by considering the near-field contributions captured by the multi-rod system and by comparing the coercivity of single- vs multi-rod systems which were described previously. We accounted for the far-field contributions by scaling coercivity with the packing fraction of the magnetic material in the non-magnetic volume. We validated this scaling through analytical magnetostatic field calculations presented in Appendix. Next, we studied the influence of the spatial arrangement of rods and of intrinsic material properties on the hysteresis properties based on the scaling of coercivity, remanence and magnetic flux density on the packing fraction. Subsequently, we extended the coercivity modelling approach from a single grain described by the multi-rod system to an ensemble of grains utilising the distribution of orientations of the grains derived from magnetic measurements of the reference sample. The results of the statistical model were used to predict hysteresis properties as a function of rod dimensions, their spatial arrangement and orientation of the grains. The results are evaluated and their significance is presented in Section 4. We summarise our findings and highlight the novelty of the work in Section 5. 

\section{Method}

\subsection{Experimental details}

The microstructure of the reference sample was investigated by Scanning Electron Microscopy (SEM). Back-scattered electrons (BSE) were detected to obtain atomic-mass contrast images of phases. Contrast between grains was observed due to orientation-dependent electron channelling effects in BSE images. We obtained the Kikuchi diffraction patterns that can be interpreted as fingerprints from given crystal structures in a given crystallographic orientation using the scanning electron microscope equipped with an electron backscatter difractometer (EBSD). Crystal orientation maps were generated, and crystal texture was depicted in inverse pole figures (IPF). \par

At the nanoscale, morpho-structural properties of the reference sample were studied by analytical high-resolution scanning transmission electron microscopy (STEM). Electron transparent specimens of a few \si{\micro m^2} in size were prepared by Focused Ion Beam (FIB). In order to study the spinodal decomposition regions, the specimens were oriented both along the growth direction (001) (longitudinal) and perpendicular to it (transverse).
The investigations were done in a FEI Titan G2 60-300 equipped with a high brightness electron gun (XFEG) with Wien-filter monochromator (\SI{200}{meV} resolution), DCOR Cs probe corrector (\SI{0.8}{\angstrom} spatial resolution in STEM) and a super-X energy dispersive spectroscope (EDS) detector (\SI{0.7}{srad} collection angle). High angle annular dark field (HAADF) was used to depict the spinodal structure with Z-contrast and EDS maps were recorded to highlight the distribution of the chemical elements.

We measured the demagnetisation curve of the reference sample via a closed-loop hysteresis graph. The magnetic field strength and the magnetic polarisation are measured simultaneously with special measuring coils. Two fluxmeters are required to process the output signals of the coils. The reference sample was saturated in the electromagnet at the start of the measurement. The measured magnetic flux was divided by the specimen area to obtain flux density and polarisation. The demagnetisation curve was measured with the sample aligned along the direction of the applied field. From the measured demagnetisation curve we extracted the remanent and saturation magnetisation that were further used in the calculation of easy axis distribution of the measured sample. 

\subsection{Micromagnetic simulations}

Micromagnetic simulations were performed to compute the demagnetisation curves of the single- and multi-rod systems described in the introduction. The magnetic material and an airbox surrounding the magnet were discretised with finite elements in order to investigate shapes and spatial variations in material properties. The role of the airbox is to effectively treat the open-boundary problem \cite{Chen1997}, when algebraic multigrid solvers are used on graphical processors \cite{Demidov2012}. Geometries are generated via the open-source computer-aided design (CAD) software Salome \cite{Salome}. Finite element meshes were generated with the software MeshGems \cite{SalomeMesh}. To ensure accurate results, the size of the finite elements was set smaller than the exchange length, $l_{ex} = \sqrt{A/(\mu_0 M_s^{2})}$, of the material, where $A$ is the exchange stiffness constant and $M_s$ is spontaneous magnetisation. The micromagnetic solver applies a preconditioned conjugate gradient method for energy minimisation \cite{Exl_CompPhysCom2019}. The Gibbs free energy includes exchange, magnetocrystalline anisotropy, magnetostatic and Zeeman energy contributions. The ramp speed of the slowly-decreasing external field is much slower than the Larmor precession to safely neglect dynamic effects \cite{Serpico}. The magnetisation configuration is computed under the constraint that the magnetisation vector remains normalised at each field step. A small field angle of \SI{1}{\degree} between the direction of the applied field and the long axes of the rods is used to avoid numerical artifacts. 

The micromagnetic simulations of the single-rod system were used to study the effect of shape and dimensions of a rod on the magnetisation reversal process, without considering interactions with neighbouring rods. The analysed rod dimensions are rod diameter and aspect ratio. Finite element micromagnetic simulations can effectively treat only a reduced number of rods. Therefore, we separated the magnetostatic interaction into a near-field contribution and a far-field contribution. The near-field interactions strongly depend on shapes and dimensions of the rods and on their spacing. These factors influence the location where the reversed domains nucleate and how the reversal propagates from one rod to another. Hence, the non-uniformity of the near-field interactions cannot be included in a simple mean-field correction and must be resolved explicitly by micromagnetics. In our modelling approach, the near-field contribution is calculated explicitly by the micromagnetic simulations of the multi-rod system, whereas the far-field contribution is generated by a larger number of rods located beyond the explicitly simulated multi-rod system. The far field acts as a mean demagnetising background field \cite{Exl2020}. The separation in near-field and far-field contributions to the dipole field is also called Lorentz sphere approximation or the effective field method. Based on the classical description of aligned elongated single-domain particle magnets, where each particle is immersed in a uniform medium of effective magnetisation, $p M_s$, parallel to the magnetisation of each particle, we represent the far-field contribution by scaling the coercive field of the multi-rod system (that accounts for the near-field contributions) with the factor $1-p$ \cite{Luborsky_JAppPhys1957, Wohlfarth_AIP1959}, where $p$ is the packing fraction. In the present work, the coercive field obtained from the simulation of a multi-rod system is therefore converted into an effective coercive field by applying the scaling factor $1-p$, with $p$ calculated based on rod dimensions and interrod spacing. Using the Lorentz sphere approximation we validated the linear dependence of coercivity on the factor $1-p$. We present the results in Appendix. 

In this work, the magnetic field strength, $H$, magnetisation, $M$, and henceforth coercivity and remanence obtained through the simulations of the multi-rod system, were scaled with $1-p$ and $p$ to account for the far-field dipolar interaction and for the amount of magnetic material within a given volume. The scaling of the magnetic field strength by $1-p$ follows from the mathematically identical differential equations governing the magnetisation rotation for both an isolated particle and an ensemble of particles \cite{Wohlfarth_ProcRSoc1955}. The maximum energy density product was obtained as the area of the largest rectangle that can be accommodated under the B-H curve, where $B = \mu_0(H + M)$. 

We extended the investigation of the magnetic properties of a single grain to those of a bulk magnet through a statistical model of grain orientation based on the angular orientation distribution of all the grains that make up the magnet. We assumed that the rods inside a given grain are well aligned with each other and that the disorder is more pronounced between grains than between rods of a grain. Previous studies based on magnetic measurements \cite{DeCampos_JOM2024} of permanent magnets have shown that the Gaussian distribution is a good approximation for the angular orientation distribution of the grains. We considered a Gaussian orientation distribution expressed as: $P (\theta ) = N \exp\left({\frac{-\theta^2}{\sigma^2}}\right)$, where $\theta$ is the angle between the easy axis of the corresponding grain and the applied field, $\sigma$ is the standard deviation and N normalises the distribution to 1. $P (\theta )$ gives the probability that a grain is misaligned by a tilt angle $\theta$. Most grains are oriented with respect to the applied field by the angle that gives the maximum of the angular orientation distribution. The standard deviation of the angular orientation distribution was derived from the normalised remanent to saturation magnetisation ratio of the reference sample. We calculated the angular orientation distribution of the grain ensemble by considering that the sample is rotated before applying a saturating field and demagnetising the sample \cite{Rieger_PhysStatSol1999}. The rotation changes the distribution function to account for the rotation angle. Grains which are strongly misoriented flip during saturating the sample in the external field \cite{Kuncser2003}. Misaligned easy axes are distributed over a sphere and are defined by their polar and azimuthal angles relative to the applied‑field direction. We integrated the modified distribution function over the azimuthal angle and subsequently evaluated it for various polar angles, $\theta$, and obtained the dependence of the modified distribution function on the tilt angle $\theta$. We estimated the coercivity of the ensemble of exchange-decoupled grains as the coercivity of a single grain tilted with respect to the applied field by the tilt angle, $\theta$, of most grains.

One of the advantages of micromagnetic modelling is the possibility to change the material properties independently of other factors involved. In this work, we used the intrinsic material properties of Fe-Co to model the $\alpha_{1}$ phase. Subsequently, we considered materials with saturation polarisation and magnetocrystalline anisotropy constant varying around those of Fe-Co in order to highlight their influence on the hysteresis properties. The intrinsic material properties used in our simulations are presented in Table \ref{tab:1}.

\begin{table}[h]
    \caption{Magnetocrystalline anisotropy constant ($K_1$), saturation polarisation ($J_s$), and exchange stiffness ($A$) of the materials used in our simulation.}
    \centering
    \label{tab:1}
    \begin{tabular}{l c c c c}
        \hline
        & $K_1$ [kJ/m$^3$] & $J_s$ [T] & $A$ [pJ/m] & Source \\
        \hline
        Fe-Co: high $J_s$, zero $K_1$
        & 0 & 2.1 & 11 & \cite{Ke_APL2017} \\

        Fe-Co$^{*}$: high $J_s$, low $K_1$
        & 15 & 2.35 & 20 & \cite{McCurrie_1982} \\

        Co: low $J_s$, zero $K_1$
        & 0 & 1.76 & 13 & \cite{Schrefl_JAppPhys1999} \\

        Co$^{*}$: low $J_s$, high $K_1$
        & 450 & 1.76 & 13 & \cite{Schrefl_JAppPhys1999} \\
        \hline
    \end{tabular}
\end{table}

Finally, in order to gain more insight into the dependence of the magnetic properties on the geometrical and intrinsic material properties of the rods and on the grain misalignment, we realised a machine-learning study. We used a multilayer perceptron regressor with 1 hidden layer consisting of 20 neurons and the rectified linear unit activation function that predicts coercivity, remanent magnetisation and maximum energy density product. To minimise the loss function during training, the limited memory Broyden–Fletcher–Goldfarb–Shanno algorithm was implemented \cite{Goodfellow_2016}. The input fed to the network consists of 4 features: diameter of the rod normalised by the exchange length of the material, aspect ratio, interrod spacing and grain misalignment. We used  800 samples for training and 200 samples for testing the performance of the model. The coefficient of determination on the test set was 0.98 for the coercivity prediction and 0.99 for the remanent magnetisation and energy density product prediction. After we confirmed that the performance of the model is high, we re-trained the model (with the same hyper-parameters) on the entire data set. We evaluated the accuracy of the trained model by analysing the residuals, defined as the difference between the true values and the corresponding model predictions.

\section{Results}
\subsection{Morpho-structural characterisation}
\begin{figure}[H]
 \centering
        \includegraphics[width=.9\textwidth]{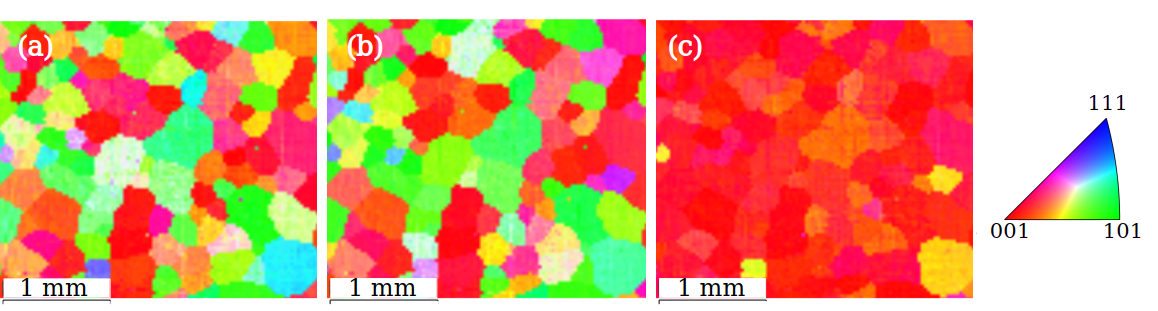}
 \caption{Inverse Pole Figure maps colored according to the crystal orientation relative to the X (a), Y (b) and Z (c) directions}
\label{fig2}
\end{figure}

TEM and IPF maps offer information about the orientation of the rods inside a grain and about the orientation of the grains with respect to a certain crystallographic direction, respectively. IPF maps coloured according to the crystal orientation relative to the X-, Y-, and Z- directions, respectively, are shown in Fig. \ref{fig2}. The IPF maps show the formation of micron-sized grains of \SIrange{200}{500}{\um} width with different orientations. This is in line with our expectation that a good magnet with a high maximum energy density product has a strong texture and large, elongated grains. 

\begin{figure}[H]
 \centering
        \includegraphics[width=0.8\textwidth]{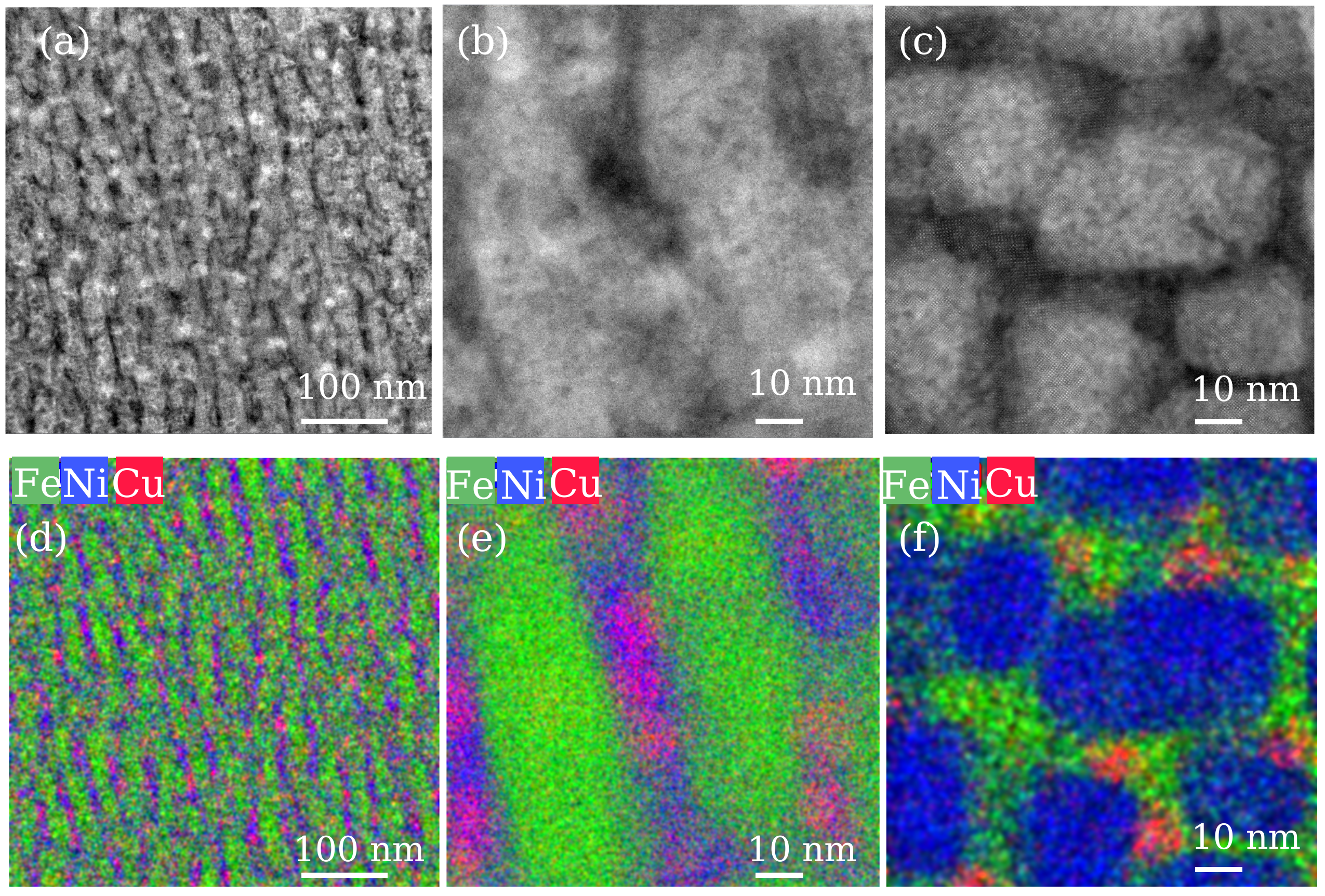}
 \caption{HAADF (a) and high-resolution HAADF (b) mappings of a longitudinal section of the rods, high-resolution-HAADF mapping of a transverse section of the rods (c). EDS (d) and high - resolution EDS (e) mappings of a longitudinal section of the rods, high - resolution EDS (f) mapping in a transverse section of the rods.}
\label{fig3}
\end{figure}

We conducted a STEM investigation of the Alnico5-7 reference sample to gain more insight into the morphology of a single grain. Fig. \ref{fig3} shows the HAADF and EDS mappings of longitudinal and transverse sections of the rods. Most of the rods have a diameter of \SI{20}{nm} and an aspect ratio of 5. Each grain contains nano-sized magnetic rods embedded in a non-magnetic matrix. The deviation from the mean orientation is much larger between grains than between rods of a grain. We present a sketch showing the disorder between grains in Fig. \ref{fig1} (c). We used the assumption of a much more pronounced disorder between grains than within a single grain in the statistical modelling of coercivity.

\subsection{Micromagnetic simulations}
\begin{figure}[H]
 \centering
        \includegraphics[width=0.98\textwidth]{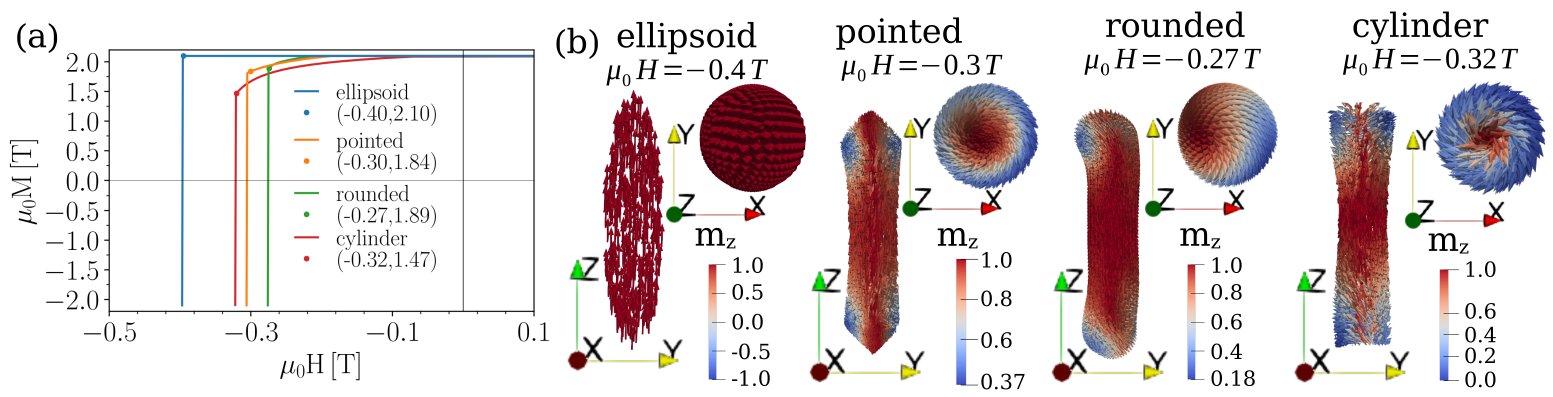}
 \caption{Demagnetisation curves of isolated ellipsoidal and cylindrical rods with flat, rounded, or pointed ends. A dot on each demagnetisation curve highlights a state preceding reversal for which we show the magnetisation configuration in subfigure (b). The width of the rod is 20 nm and the aspect ratio is 5 (a). The corresponding magnetisation configurations observed in longitudinal cross section and transverse projection in an applied field that precedes reversal. The rod shape and the applied field in which magnetisation configuration was registered, are indicated on top of each plot. The colour bar represents the vertical component of the magnetisation vector, $m_z$ (b). }
\label{fig4}
\end{figure}

In an ideal Alnico magnet the rods are exchange-decoupled to allow magnetisation to be spatially confined within individual nano-sized elements. Isolation suppresses the propagation of reversed domains into adjacent rods. Consequently, the system would require a higher applied field to switch the magnetisation compared to configurations in which exchange coupling facilitates switching \cite{Ke_APL2017}. We first examined individual rods with different shapes by varying the roundness of their ends. We considered ellipsoidal-shaped rods and cylindrical rods with flat, rounded or pointed ends. The simulated demagnetisation curves of an isolated Fe-Co rod, with dimensions identified via TEM and various rod shapes are shown in Fig. \ref{fig4} together with magnetisation states preceding reversal, observed in longitudinal and transverse projections. The examined magnetisation states of the reversal are marked with dots on the corresponding demagnetisation curves. 

In systems with reduced magnetocrystalline anisotropy - such as bcc Fe-Co - the shape anisotropy is the only mechanism that contributes to improving coercivity \cite{Luborsky_JAppPhys1957, Chikazumi_PhysMag1964}. The magnetisation of rods with dominant shape anisotropy energy has two stable minima with magnetisation parallel to the long axis of the rod. Magnetic surface charges generate a strong demagnetising field that locally reduces the magnetostatic energy and facilitates the nucleation of reversed domains at the ends of the rod \cite{Givord_JPhysIV1992}. The nucleation of the reversed domain was studied as an energy barrier that is decreased by an increasing external field. Once a reversed domain nucleates, its propagation throughout the rod volume leads to magnetisation reversal in a single step \cite{Forster_JMMM2003}. 

How coercivity is influenced by shape depends on rod dimensions and on the presence of magnetocrystalline anisotropy as a material property. We discuss the reversal process for rods with aspect ratio of 5, specific to the reference sample. Micromagnetic simulation results for rods with diameters of \SI{20}{nm}, as observed in the reference sample, are presented in Fig. \ref{fig4}. We also discuss the magnetisation reversal process for rods with aspect ratio of 5 and diameters below and above \SI{20}{nm} (not shown). Our micromagnetic simulations show that for rods prepared from materials with low magnetocrystalline anisotropy, and with diameters below \SI{20}{nm} for which the relative importance of the exchange energy to the magnetostatic energy is high, more rounded rod ends reduce surface charges and the resulting decrease in magnetostatic energy limits nucleation. Reversal occurs by formation of reversed domains at the ends of the rods. Due to the more homogeneous demagnetising field, increasing the roundness of the rod ends from cylindrical to ellipsoidal-shaped rods, leads to an increase in coercivity. For rods with diameters of \SI{20}{nm} and above, the relative importance of exchange to magnetostatic energy decreases. The simulations give evidence of vortex formation at the ends of the rods depending on the shape and dimensions of the rod. For ellipsoidal-shaped rods with diameter of \SI{20}{nm}, uniform magnetisation persists up to a reverse field of \SI{0.4}{T}. For \SI{20}{nm} thick rods, all rod shapes except the ellipsoids show vortices formation at the ends. In the case of rods with rounded ends, the vortices tilt with decreasing the applied field and magnetisation is reversed in the entire volume at an applied field whose magnitude is lower than that required to reverse magnetisation in rods with flat ends. As a consequence, the coercivity of rods with rounded ends is lower than that of rods with flat ends. For rods with pointed and flat ends, outer-shell spins curl around the long axis of the rods starting with the ends and subsequently propagating through the volume upon increasing the applied field. Central spins, located around the long axis of the rod, reverse in a higher applied field than outer-shell spins. The same behaviour is observed for thicker rods, regardless of their shape. Our micromagnetic simulations showed a similar coercivity of elipsoidal and pointed-ended rods and of cylindrical and rounded-ended rods for rods with diameter of \SI{30}{nm}. For rods with diameter of 40 and \SI{50}{nm}, the coercivity of ellipsoidal rods is higher than that of the other shapes. The coercivity of rods with these thicknesses is weakly dependent on shape for rods with reduced roundness of the ends.

The magnetocrystalline anisotropy promotes uniform magnetisation at the ends of the rod and makes formation of reversed domains dependent on the homogeneity of the demagnetisation field at the ends independently of rod dimensions \cite{Kuncser2014}. The uniformity of the magnetostatic field in rods fabricated from materials with high uniaxial magnetocrystalline anisotropy determines magnetisation reversal through formation of reversed domains at the rod ends and subsequent propagation in the volume. This type of reversal leads to higher demagnetisation curve squareness and coercivity with increasing roundness of rod ends due to the shape-dependent uniformity of the demagnetising field \cite{Bance_JMMM2014, Fischbacher_JPhysD2018}.

\begin{figure}[H]
 \centering
        \includegraphics[width=0.98\textwidth]{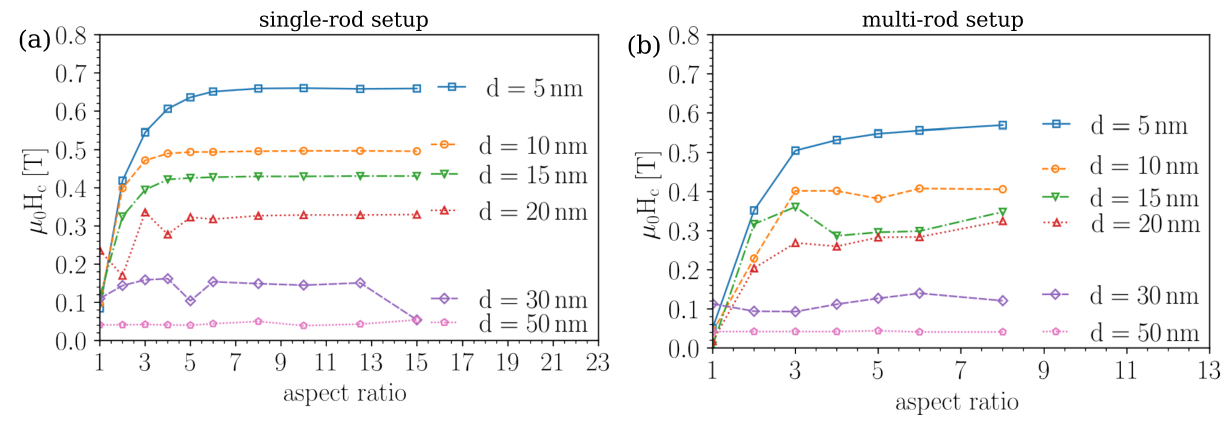}
 \caption{Coercivity as a function of aspect ratio for different rod diameter for the single- (a) and multi- (b) rod systems. The lateral and vertical interrod spacing in the case of the multi-rod system is 5 nm.}
\label{fig5}
\end{figure}

In the modelling of a single grain, we assumed the rods to be cylindrical. We compared the coercivity of single- and multi-rod systems with rod dimensions varying across those observed experimentally. We analysed diameters ranging from 5 to $\SI{50}{nm}$ and aspect ratios ranging from 1 to 15 for the single-rod system and from 1 to 8 for the multi-rod system. In Fig. \ref{fig5} (a) and (b) we present the coercivity of Fe-Co single and multi-rod systems as a function of the aspect ratio for different diameters of a rod. 

Changing the size of the magnetic sample modifies the relative weight of the different energy terms towards the total Gibbs free energy. Increasing the shape anisotropy, either by decreasing the diameter or by increasing the length of the rod, increases coercivity. At a fixed length, decreasing the diameter of the rod increases the relative importance of the exchange energy to magnetostatic energy. The magnetisation remains uniform in stronger opposing external fields and the formation of reversed domains is suppressed \cite{Exl2020}. Consequently, a higher external field is needed to switch magnetisation, leading to an increase in coercivity. The coercivity shows a strong dependence on the aspect ratio below 8 for diameters up to 20 nm. Above the aspect ratio of 8 no significant improvement of the coercivity can be observed. This plateau in coercivity at higher aspect ratios occurs because the longitudinal demagnetisation factor of the cylindrical nanorods decays asymptotically, causing the purely geometric shape anisotropy contribution to reach its theoretical maximum limit. The threshold value of the aspect ratio above which coercivity remains nearly constant can be attributed to the diminishing influence of the reversed demagnetised regions relative to the total rod volume. The dependence of coercivity on aspect ratio becomes weaker with increasing rod diameter above \SI{20}{nm}. This observation can be explained by the associated diminished relative importance of the exchange towards magnetostatic energy that is less dependent on aspect ratio for thick diameters. The expansion of the vortices from the rod ends throughout the rod volume occurs at a lower magnitude of the applied field as the rod diameter increases, or as the aspect ratio decreases. The shape of the demagnetisation curves (not shown) gives evidence that the propagation of the vortices in high aspect ratio rods is slower than in low aspect ratio rods for both \SI{20}{nm} and \SI{50}{nm} thick rods. However, as the diameter is increased, the applied field necessary to cause the formation of a vortex throughout the rod becomes less dependent on aspect ratio. Intermediate states of the reversal consisting of vortex formation throughout the rod are formed by the curling of outer-shell spins around the long axis of the rod over the entire length, whereas central spins are still predominantly oriented in the direction of the saturation field. For \SI{50}{nm} thick rods, this field is independent of aspect ratio leading to the same effect in coercivity. Simulations of the multi-rod system have been carried out only for aspect ratios up to the threshold of 8. 
Magnetostatic interactions between rods reduce the coercivity with respect to that of a single-rod system. The stray field generated by a rod adds up to the demagnetising field of its neighbours. Hence, the magnetisation reversal mechanism of multi-rod systems differs from that of isolated rods by the stray field each rod generates on its neighbours. The magnetostatic field distribution, including both the demagnetising and stray fields, governs the magnetisation reversal process of multi-rod systems. For example, the coercivity of an Fe-Co multi-rod system with a rod diameter of $\SI{20}{nm}$, a height of $\SI{100}{nm}$ and an interrod spacing of $\SI{5}{nm}$, is reduced by approximately $50 \%$ with respect to that of a single-rod. The dependence of coercivity on rod dimensions follows the same trends for both single- and multi-rod systems.

\begin{figure}[H]
 \centering
        \includegraphics[width=0.48\textwidth]{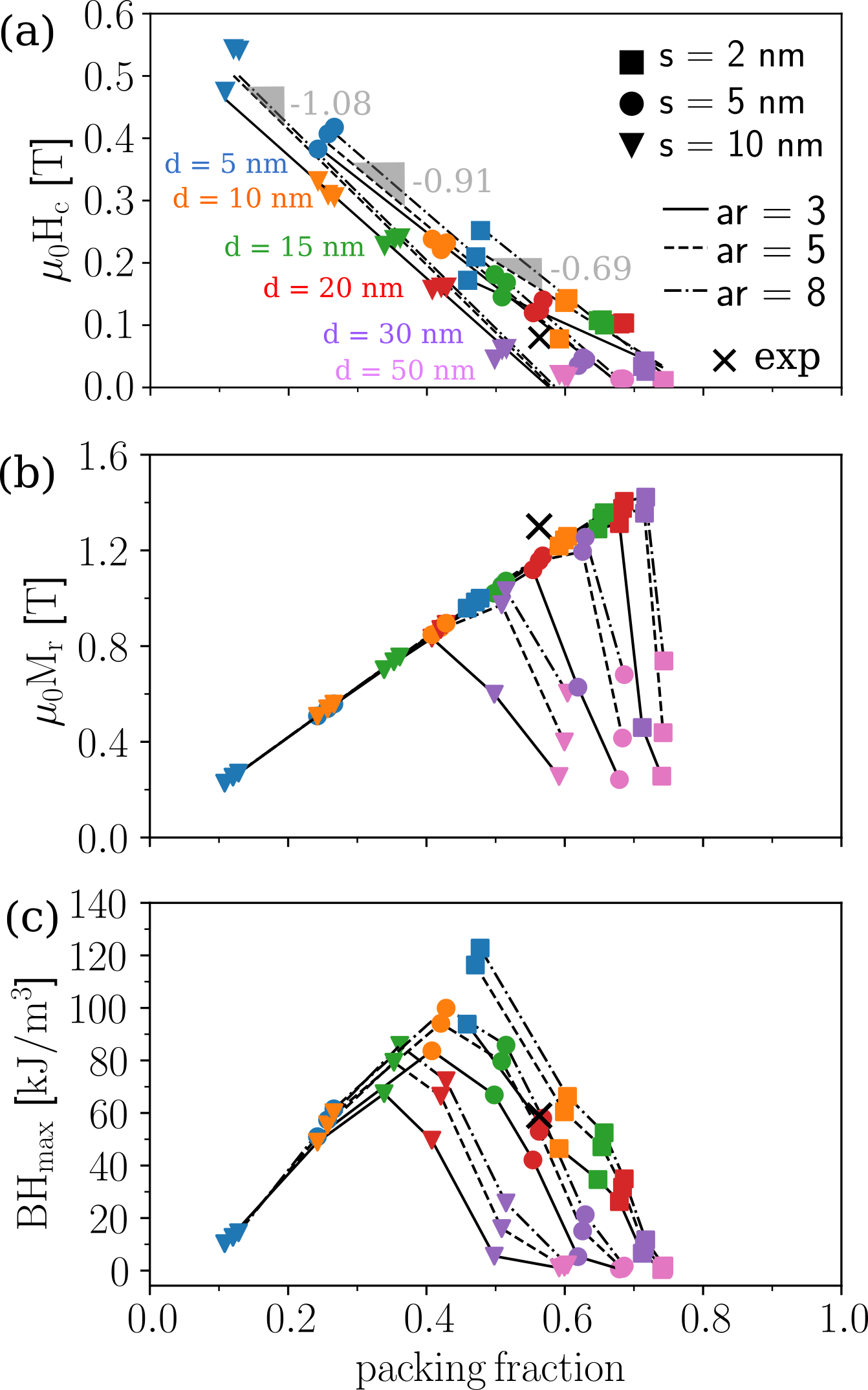}
 \caption{Coercivity (a), remanent magnetisation (b) and maximum energy density product (c) as functions of packing fraction, $p$. Rod diameters, $d$ and interrod spacings, $s$ are indicated by different colours and markers, respectively. Vertical end-to-end spacing is 5 nm. In (a), the linear fit of coercivity versus $p$ is shown with different line styles for each aspect ratio. The slopes of the linear fit of coercivity versus packing fraction for the $s$ considered and an aspect ratio of 5 are shown in the plot. The magnetic properties of the reference sample are indicated by an $\times$ in each plot at the corresponding packing fraction computed based on the geometric parameters identified by TEM: $\mathrm{d = 20 \ nm}$, $\mathrm{ar = 5}$ and $\mathrm{s = 5 \ nm}$.}
\label{fig6}
\end{figure}

Besides the rod dimensions, the magnetic properties of the multi-rod system also depend on the interrod spacing. Fig. \ref{fig6} shows the coercivity, remanent magnetisation, and maximum energy density product as functions of the packing fraction. We compared the simulation results with the coercivity, remanent magnetisation and maximum energy density product of the reference sample. The experimental values are indicated in the plots by an $\times$ at the corresponding packing fraction, determined from the rod dimensions and interrod spacing of the reference sample. The good agreement between the measured and simulated values validates the simulation methodology. 

Increasing packing fraction, either by increasing rod diameter or by decreasing interrod spacing, diminishes the coercivity. The nucleation of reversed domains or vortex formation at rods ends is facilitated by the increase of the rod diameter, leading to a corresponding decrease in coercivity. As the interrod spacing decreases, each rod is subjected to a stronger stray field generated by its adjacent neighbours, which adds to the applied field and consequently reduces the coercivity. We performed an ordinary least squares linear regression of the dependence of coercivity on packing fraction for fixed interrod spacing and aspect ratio. The slopes derived from the linear fits corresponding to an aspect ratio of 5 are indicated in the plot. The stronger magnetostatic interactions make the dependence of the coercivity on packing fraction more sensitive to changes in diameter and aspect ratios as the interrod spacing decreases. The remanent magnetisation increases with increasing packing fraction due to an increase in the amount of magnetic material within a given volume. However, for diameters exceeding \SI{30}{nm}, the remanent magnetisation is reduced as a result of earlier magnetisation reversal induced by the increased demagnetising field at the rod ends. Optimisation of the maximum energy density product necessitates balancing coercivity and remanent magnetisation. The simulation results suggest that the maximum energy density product is favoured by thin rods (ideally with diameters below \SI{20}{nm}) and high aspect ratios (but not exceeding a threshold of 6). Large interrod spacing promotes high coercivity. However, the increase in interrod spacing should be constrained by the need to maintain a high volume fraction of magnetic material, which is required to achieve substantial remanent magnetisation. For an optimal maximum energy density product, the interrod spacing should be adjusted according to the rod diameter.

\begin{figure}[H]
 \centering
        \includegraphics[width=0.48\textwidth]{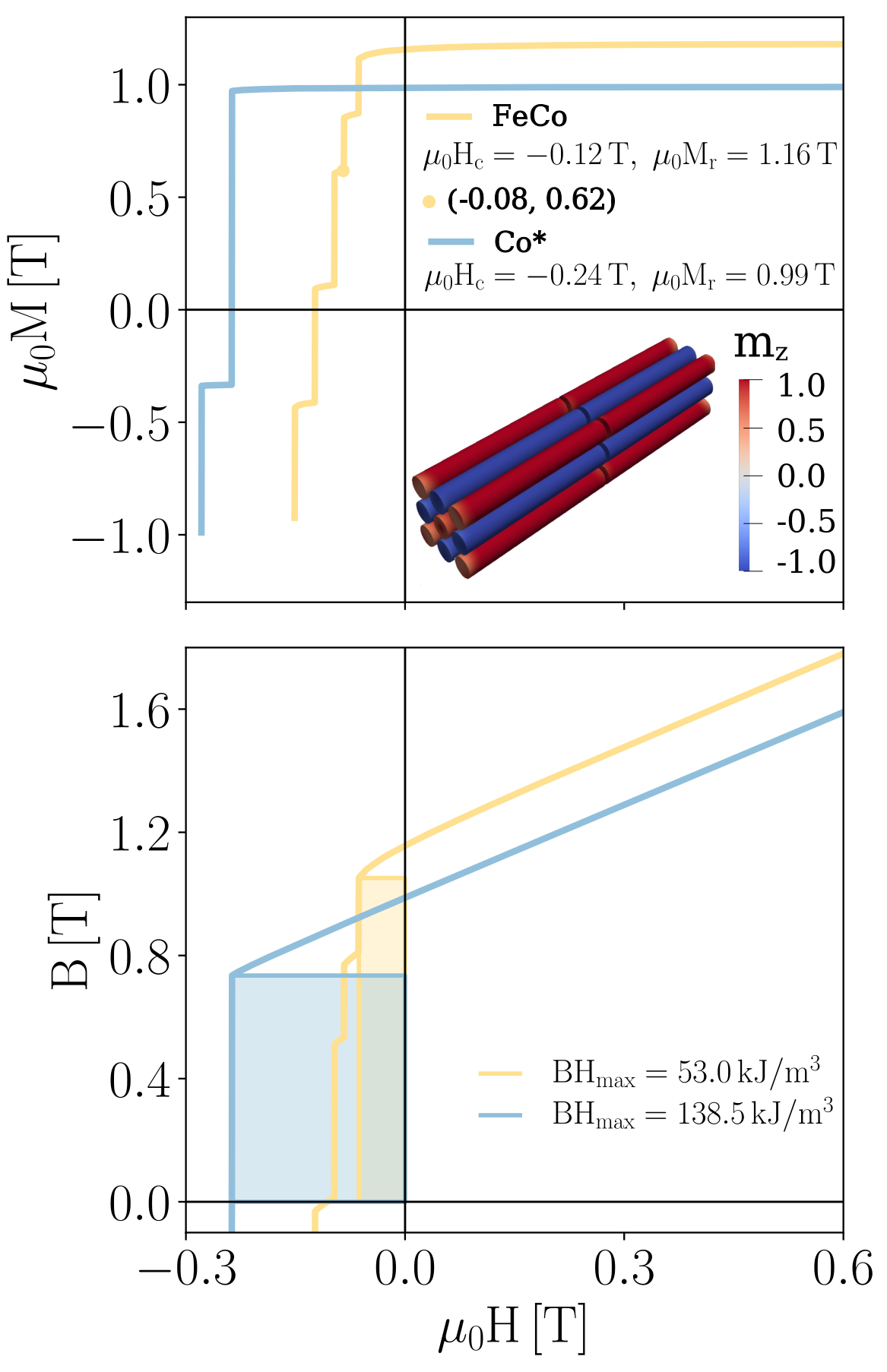}
 \caption{Demagnetisation curves of Fe-Co and Co$^*$ multi-rod systems and map of the $z$ component of the magnetisation in an applied field of -\SI{0.08}{T}. The coordinates of the point on the demagnetisation curve corresponding to the investigated magnetisation configuration of the Fe-Co system is shown in inset together with the hysteresis properties of both Fe-Co and Co$^*$ (a). $B$-$H$ curves computed based on the demagnetisation curves (b). The diameter of the rods is $\SI{20}{nm}$ and their aspect ratio is 5. The interrod and vertical end-to-end spacing is $\SI{5}{nm}$. The coordinate system indicates the orientation of the rods, with their long axis aligned parallel to the z-axis.}
\label{fig7}
\end{figure}

We investigated the magnetisation reversal process of Fe-Co and Co$^*$ multi-rod systems with dimensions of the rods and interrod spacing similar to those evidenced by the TEM analysis of the reference sample. By comparing Fe-Co and Co$^*$ we studied the effect of zero (Fe-Co) or high (Co$*$) magnetocrystalline anisotropy constant and of low (Co$^*$) or high (Fe-Co) saturation polarisation. We present the demagnetisation curves of the Fe-Co and Co$^*$ multi-rod systems and the $B$-$H$ loops computed based on the demagnetisation curves in Fig. \ref{fig7}. The map of the z-component of magnetisation of the Fe-Co multi-rod system in an applied field of -\SI{0.08}{T} is shown in the inset of Fig. \ref{fig7} (a).  

The demagnetisation curves show steps at applied fields corresponding to magnetisation reversal in one or several rods. The map of the component ($m_z$) of magnetisation along the long axis of the rods in a field of \SI{80}{mT} supports this observation. In multi-rod systems with rod diameter of \SI{20}{nm} and aspect ratio 5, the z-component of magnetisation, $m_z$, deviates from 1 at the ends of the rods in incipient states of the reversal as a consequence of the inhomogeneous stray fields around the edges that promote the nucleation of reversed domains. Increasing the magnitude of the applied field, the reversed domains propagate in the rod volume and the magnetisation in the rod is switched. The switching of all rods depends on the interplay between the applied field that opposes the initial magnetisation and the stray field of already switched rods that contribute to the stabilisation of the magnetisation of their neighbours against switching. In a partially reversed state the magnetisation of some of the rods is switched. A further increase of the applied field switches the magnetisation of all or almost all rods. 
The higher field required to overcome the total magnetic anisotropy energy barrier in Co$^*$, the material that possesses high magnetocrystalline anisotropy, improves coercivity. The coercivity is approximately \SI{0.12}{T} for Fe-Co and \SI{0.24}{T} for Co$^*$. The remanent magnetisation is \SI{1.16}{T} for Fe-Co and \SI{0.99}{T} for Co$^*$. The higher remanent magnetisation of Fe-Co as compared to that of Co$^*$ is attributed to its higher saturation magnetisation. The estimated maximum energy density product is 53.2 and 136.9 $\mathrm{kJ/m^3}$ for Fe-Co and Co$^*$, respectively. 

\begin{figure}[H]
 \centering
        \includegraphics[width=0.48\textwidth]{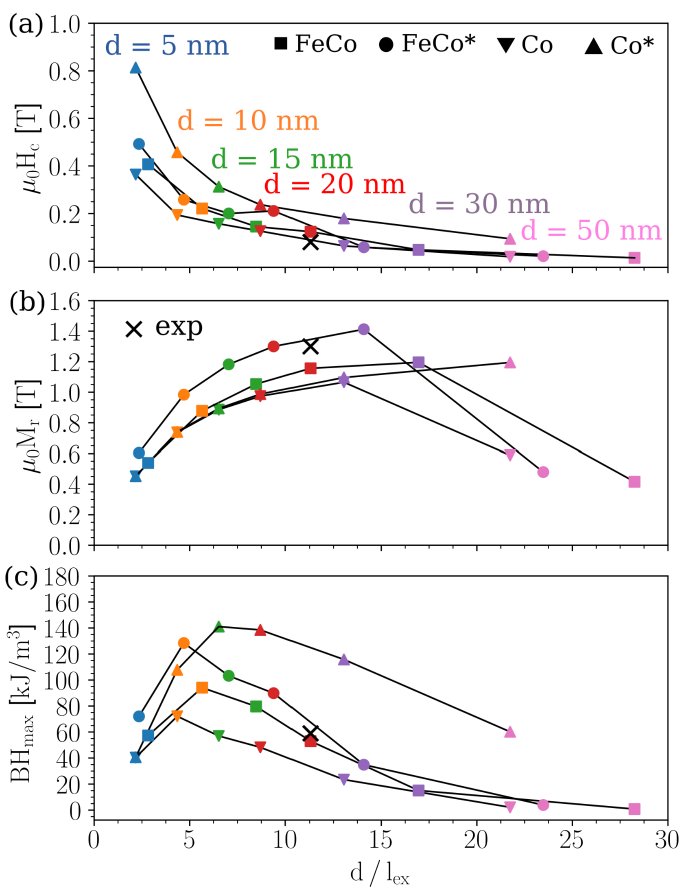}
 \caption{Coercivity (a), remanent magnetisation (b) and maximum energy density product (c) as functions of normalised parameter, $d/l_{ex}$. Different materials are represented by distinct marker shapes, whereas diameter values are differentiated by color. The aspect ratio considered is 5.  The interrod and the vertical end-to-end spacing are set to $\SI{5}{nm}$. The coercivity, remanence and maximum energy density product of the reference sample are indicated by an $\times$ in each plot.}
\label{fig8}
\end{figure}

We analysed the impact of intrinsic material properties on the hysteresis properties for various rod diameters. The aspect ratio considered is 5, in line with the experimental observation of the reference sample. Fig. \ref{fig8} presents the coercivity, remanent magnetisation and maximum energy density product for different materials and rod diameters as function of a dimensionless parameter defined as diameter of the rod divided by the exchange length of the corresponding material. The results presented here are obtained under the assumption of an ideal arrangement of exchange-decoupled rods, with all rods oriented along the direction of the applied field. 

For a fixed saturation polarisation, the high uniaxial magnetocrystalline anisotropy constant of \SI{450}{kJ/m^3} of Co$^*$ doubles coercivity for the entire range of investigated diameters. A low magnetocrystalline anisotropy constant of \SI{15}{kJ/m^3} of Fe-Co$^*$ improves coercivity with respect to materials with zero uniaxial magnetocrystalline anisotropy constant. For example, coercivity increases by approx. \SI{100}{mT} and by \SI{40}{mT} for \SI{5}{nm} and \SI{20}{nm} thick rods, respectively, if Fe-Co$^*$ vs Fe-Co and Co are compared. The improvement is less pronounced for thicker rods. Fe-Co$^*$, characterised by a high saturation polarisation of \SI{2.35}{T}, shows the highest remanent magnetisation among the considered materials over the entire range of investigated diameters. The highest remanence of \SI{1.4}{T} is obtained for \SI{30}{nm} thick rods. With a saturation polarisation of \SI{2.1}{T}, specific to Fe-Co, a remanent magnetisation of \SI{1.18}{T} is obtained for \SI{30}{nm} thick rods. The remanence of Fe-Co systems is higher than that of Co and Co$^*$ for diameters up to \SI{30}{nm}, the improvement in remanence being of \SI{170}{mT} and \SI{100}{mT} for \SI{10}{nm} and \SI{30}{nm} thick rods, respectively. For \SI{50}{nm} thick rods, the uniaxial magnetocrystalline anisotropy of Co$^*$ that promotes the alignment of the magnetic moments towards the easy axis - which is the same with the direction of the applied field - overpasses the effect of the reduction of remanent magnetisation with increasing rod diameter. For a rod diameter of \SI{50}{nm}, the remanence of Co$^*$ exceeds that of the rest of the investigated materials. The highest $\BHmax$ of \SI{145}{kJ/m^3} observed with our micromagnetic simulations is that of Co$^*$ \SI{15}{nm} thick rods due to the high uniaxial anisotropy constant. The $\BHmax$ of Co$*$ systems is higher than that of the rest of the investigated materials for rod diameters ranging from \SI{15}{nm} to \SI{50}{nm} where a $\BHmax$ of \SI{60}{kJ/m^3} is obtained. With a low magnetocrystalline anisotropy constant and a high saturation polarisation specific to Fe-Co$^*$, we obtained a maximum energy density product of \SI{135}{kJ/m^3} for \SI{15}{nm} thick rods, representing the second best $\BHmax$ value obtained. For fixed rod diameters of 5 and \SI{10}{nm}, the high remanence of Fe-Co$^*$ determines a $\BHmax$ slightly higher than that of Co$^*$ systems. In the absence of uniaxial magnetocrystalline anisotropy, the maximum energy density product benefits more from moderate (Fe-Co) than from low (Co) saturation polarisation. 

\subsection{Statistical treatment of the ensemble of grains by orientation}
\begin{figure}[H]
 \centering
        \includegraphics[width=0.98\textwidth]{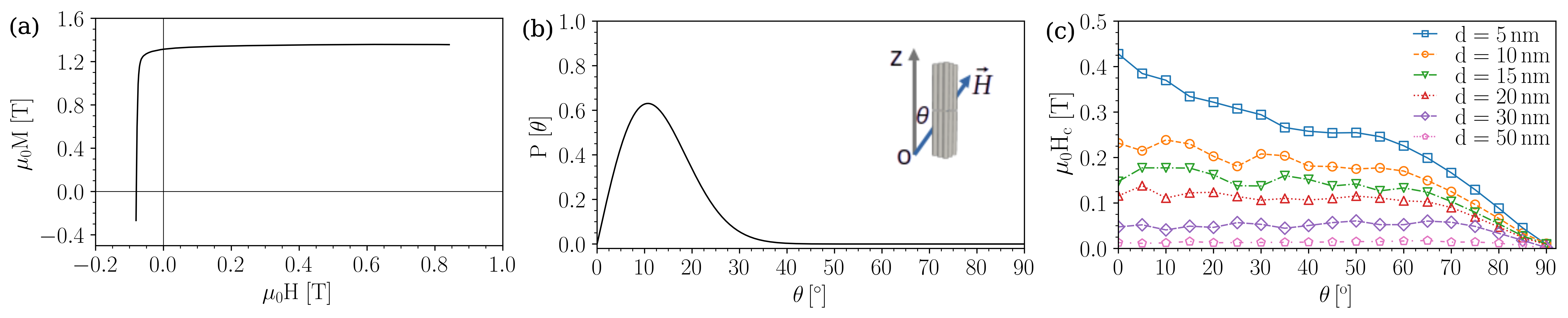}
 \caption{Measured demagnetisation curve of the reference sample (a). Distribution of easy‑axis orientations as determined from magnetic characterisation of the reference sample. The inset illustrates the measurement geometry, showing the tilt angle $\theta$ of the easy axis relative to the applied field. (b) Angular dependence of the coercivity for systems of rods with different diameters. Rod diameters are indicated by different colors. The interrod and vertical end-to-end spacings are $\SI{5}{nm}$ (c).}
\label{fig9}
\end{figure}

We described the coercivity of the ensemble of grains by implementing a statistical model based on grain orientations, as described in section Method. The demagnetisation curve used to compute the distribution of orientations of the grains is presented in Fig. \ref{fig9} (a). The computed distribution of orientations, $P\left(\theta\right)$, is presented in Fig. \ref{fig9} (b). It describes the number of grains that are tilted by an angle $\theta$ with respect to the applied field. Fig. \ref{fig9} (c) shows the coercivity of a single grain tilted by the angle $\theta$ with respect to the alignment direction. We considered grains of rods with different diameters. The rods in a grain have the same diameter. A grain consisting of rods with diameter below \SI{20}{nm} shows angular dependence of the coercivity consistent with the Stoner–Wohlfarth model. The coercivity becomes less dependent on the angle of the applied field for rod diameters higher than \SI{30}{nm}. This observation can be explained by a coercivity less dependent on the measurement geometry in the case of magnetisation reversal through propagation of vortices. The maximum of the distribution $P (\theta)$ occurs at a tilt angle of approximately $\mathrm{10^{\circ}}$, indicating that the majority of the grains are oriented by this angle relative to the direction of the applied field. We estimated the coercivity of the ensemble of grains as the coercivity of a single grain tilted by $\mathrm{10^\circ}$ with respect to the applied field.
 
Using the geometric parameters derived from the TEM analysis of the reference sample and the intrinsic material properties of Fe-Co, the coercivity computed for the ensemble of grains is \SI{110}{mT}. The coercivity of the reference sample is \SI{80}{mT}.

\begin{figure}[H]
 \centering
        \includegraphics[width=0.98\textwidth]{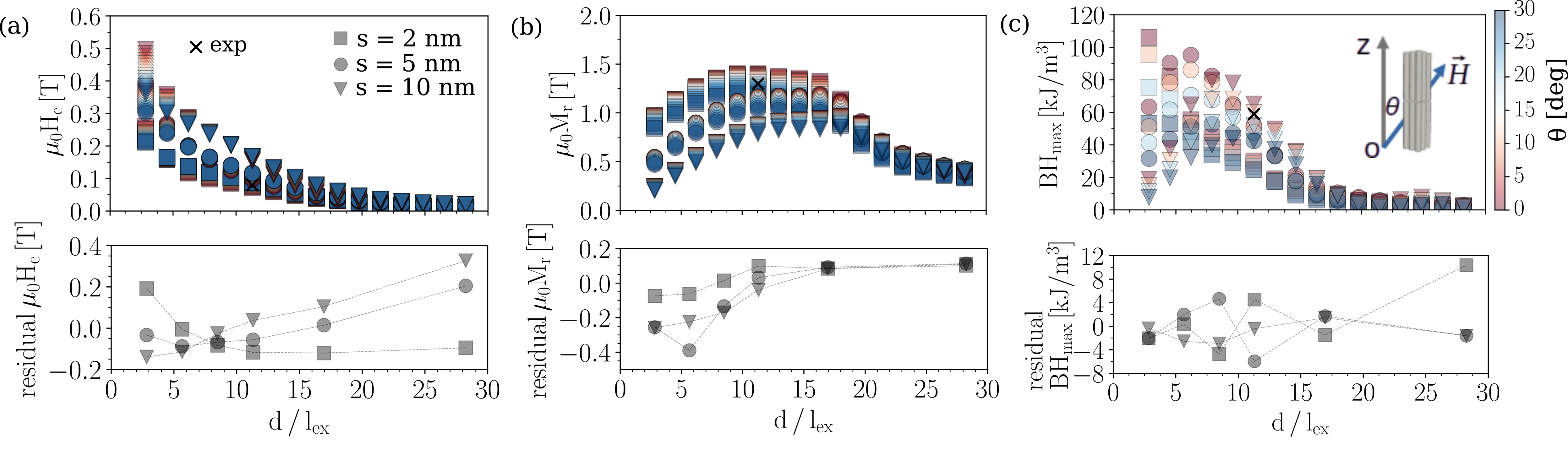}
 \caption{Coercivity (a), remanent magnetisation (b), maximum energy density product (c) and the corresponding residuals as functions of rod diameter for different interrod spacing, s, degrees of misalignment and aspect ratio 5. The experimental value is indicated by an $\times$ in each plot. Because residuals depend on the training data points, the number of residual data points is smaller than the number of predicted data points.}
\label{fig10}
\end{figure}

Based on the knowledge gained from micromagnetic simulation results, we further trained a shallow neural network to obtain fine-grained, interpolated results. The predicted coercivity, remanent magnetisation, maximum energy‑density product and the corresponding residuals are presented as functions of the rod diameter normalised by the exchange length of Fe-Co in Fig. \ref{fig10}. We considered various degrees of grain misalignment. The residual distribution is centred around zero, indicating that the model captures the dominant trends in the data. A small number of residuals fall outside the main cluster. These outliers suggest that the model struggles near the boundaries of the input space. These deviations arise from insufficient representation of these fringe regimes in the training data.

Being trained on the micromagnetic simulations results, the same trends in the dependence of the hysteresis properties on rod dimensions and interrod spacing are observed in the values predicted by the neural network. Increasing the interrod spacing leads to an enhancement in coercivity by diminishing the strength of magnetostatic interactions between neighbouring rods. The increase of the interrod spacing decreases remanent magnetisation because of the reduced amount of magnetic material within a given volume. The misalignment leads to a reduction in both the coercive field and the remanent magnetisation, which is attributed to the diminished projection of the magnetisation along the measurement direction. Consequently, the maximum energy density product is lowered. Under the assumption of perfect grain alignment, the attainable maximum energy‑density product is governed by the competition between achieving high coercivity and maintaining sufficient remanence. A high maximum energy density product is achieved when both coercivity and remanent magnetisation are relatively large, the ideal interrod spacing being dependent on the diameter of the rod.

The predicted hysteresis properties values for the grain ensemble show evidence for geometric parameters and grain misalignment characteristic to the reference sample. For example, configurations such as: rod diameter of \SI{17}{nm}, interrod spacing of \SI{2}{nm} and grain misalignment less than $\mathrm{15^{\circ}}$, rod diameter of \SI{20}{nm}, interrod spacing of \SI{5}{nm} and perfectly aligned grains or rod diameter of \SI{23}{nm}, interrod spacing of \SI{5}{nm} and grain misalignment of approx. $\mathrm{15^{\circ}}$ could explain the coercivity of the reference sample. The remanence of the reference sample of \SI{1.31}{T} is specific to rod diameter in range 14 to \SI{29}{nm}, interrod spacing of \SI{2}{nm} and grain misalignment from $\mathrm{15^{\circ}}$ to $\mathrm{30^{\circ}}$. The configurations that support the $\BHmax$ of the reference sample of \SI{59}{kJ/m^3} are rod diameter less than \SI{20}{nm}, interrod spacing of \SI{5}{nm} and grain misalignment dependent on rod diameter, rod diameter of \SI{20}{nm}, interrod spacing of \SI{10}{nm} and grain misalignment in the range $\mathrm{10^{\circ}}$ to $\mathrm{20^{\circ}}$, or rod diameter of \SI{20}{nm}, interrod spacing of \SI{5}{nm} and perfectly aligned grains. For rod diameter, interrod spacing and aspect ratio observed in the reference sample, the relative error in the computed $\BHmax$ compared to the experimental value is \SI{6.7}{\%}. 

\section{Discussion}
Previous works have treated the magnetic properties of nano-sized rod arrays for their possible applications by considering materials with high (Co) or zero (Fe) uniaxial magnetocrystalline anisotropy \cite{Ener2018, Elmekawy2021}. It has been shown that in the absence of uniaxial magnetocrystalline anisotropy, the shape anisotropy and magnetostatic interactions dictate the magnetic properties of the rod arrays \cite{Bance_JMMM2014, Kuncser2017}. Therefore, the factors that influence the magnetic properties in this case are the shape and dimensions of the rod and the interrod spacing \cite{Bance_JMMM2014, Ke_APL2017}. These factors define the packing fraction of the magnetic rods in the non-magnetic matrix. The dependence of coercivity on the packing fraction has been studied in both theoretical and experimental works and the optimal packing fraction of 0.7 was identified \cite{Panagiotopoulos2013}. At the millimeter to micrometer scale, the magnet is composed of grains with various orientations \cite{Zhou_ActaMaterialia2014}. In the present work, we realised a systematic study of the magnetic properties of Alnico magnets based on intrinsic material properties, geometric parameters of the rods and grain orientations. We progressed from a micromagnetic study of the shape and dimensions of a single-rod towards the study of rod arrays taking into account near- and far-field interactions and finally to an ensemble of millimeter-to-micrometer-sized grains treated statistically based on the distribution of orientations of the grains.  We used micromagnetic simulations of a small number of rods to account for the near-field interactions, whereas far-field contributions were incorporated based on the classical description of aligned elongated single-domain particle magnets immersed in a uniform medium of effective magnetisation. 

Our micromagnetic simulations evidence the structural parameters of the rods, intrinsic material properties and grain alignment that should be targeted for an optimal maximum energy density product of Alnico magnets. The maximum energy density product depends on both coercivity and remanence. Small rod diameters (with a linear dependence of coercivity on packing fraction), high aspect ratio (up to a threshold) and large interrod spacing improve coercivity. However, increasing the interrod spacing lowers remanence. As a consequence, the ideal interrod spacing depends on rod diameter.
If the preparation of rods from materials that possess uniaxial magnetocrystalline anisotropy is experimentally achievable, this could double the coercivity. A slight increase in the saturation polarisation and uniaxial magnetocrystalline anistropy constant could improve the maximum energy density product with up to \SI{40}{kJ/m^3} for \SI{10}{nm} thick rods. For rods with diameter below \SI{20}{nm}, the dependence of coercivity on grain alignment is more sensitive than for thicker rods. Grains composed of thin rods should be oriented towards the direction of the applied field to maintain a good coercivity.  

\section{Conclusions}

We realised a micromagnetic study of the influence of the shape and dimensions of rods, intrinsic material properties and grain alignment on the magnetic properties of Alnico magnets. We designed the geometric parameters of the rods and considered grain orientation according to those of a reference sample. Finally, we applied machine learning techniques to refine the micromagnetic simulations data and to derive tailored nanostructure and grain alignment that would improve the magnetic properties of the reference sample. We predicted coercivity, remanent magnetisation and maximum energy density product as a function of the spontaneous magnetisation, rod diameter, aspect ratio, and misalign angle of the grains using a multi-layer Perceptron regressor. For Fe-Co rods with diameters below \SI{20}{nm}, the increased roundness of rod ends improves coercivity. A further increase in rod diameter determines the formation of vortices at the rod ends in the incipient stages of the reversal. In this case, the coercivity of rods with flat ends exceeds that of pointed and rounded ends. The coercivity of shape-anisotropy based Alnico can be improved by tailoring dimensions of the rods to minimise the demagnetisation factor along their long axis. Rods with small diameters and high aspect ratio (up to a threshold of 8) favour high coercive field. Large interrod spacings improve coercive field by diminishing the magnetostatic interactions. However, increasing the interrod spacing is detrimental to remanence because of the resulting lower amount of magnetic material within a given volume. As a result of the interplay of coercivity and remanence for a good maximum energy density product, the interrod spacing should be adjusted as a function of the diameter of the rods.
If experimentally achievable, a material with high uniaxial magnetocrystalline anisotropy could double the coercivity and maximum energy‑density product. A slight increase of the uniaxial magnetocrystalline anisotropy and of the saturation polarisation improves the maximum energy density product. In the absence of uniaxial magnetocrystalline anisotropy, the maximum energy density product is enhanced by the saturation polarisation. The hysteresis properties can be improved with respect to those of the reference sample by rods with a diameter below $\SI{20}{nm}$ and tailored interrod spacings. The misalignment of the grains lowers the performance of the magnet because of the diminished projection of the magnetisation along the measurement direction. The stronger shape anisotropy of rods thinner than \SI{20}{nm} makes the magnetic figures of merit more sensitive to misalignment than those of thicker rods.

\section*{Acknowledgements}
Funding by the European Union within the MagNEO project (grant No. 101130095) is acknowledged. Views and opinions expressed are however those of the authors only and do not necessarily reflect those of the European Union or the European Health and Digital Executive Agency (HADEA). Neither the European Union nor the granting authority can be held responsible for them. 

The authors declare no conflicts of interest.

\section*{Data Availability}
Micromagnetic simulations data that support the findings of this study has been made available at: \url{https://zenodo.org/records/19898175?preview=1&token=eyJhbGciOiJIUzUxMiJ9.eyJpZCI6ImIwYjg1Nzc3LTBlZDUtNGFlOS1hMmI4LWFmMDAwNjhhYmMxYyIsImRhdGEiOnt9LCJyYW5kb20iOiI0YzM0ZDk4ZDgxZDMxZGRhMTgwYTQwZDY0NWQyNzViNSJ9.DsgjmGndOSfEt25coH7w3T1A9uAYSFhh02zCgLCalhSLA3fpuczRvqCZAUN6_TZXIK-gsyUurhdrkmO0w2v3qw}.




\appendix
\section{Analytical computation of the stray field}
\begin{figure}[H]
 \centering
        \includegraphics[width=0.98\textwidth]{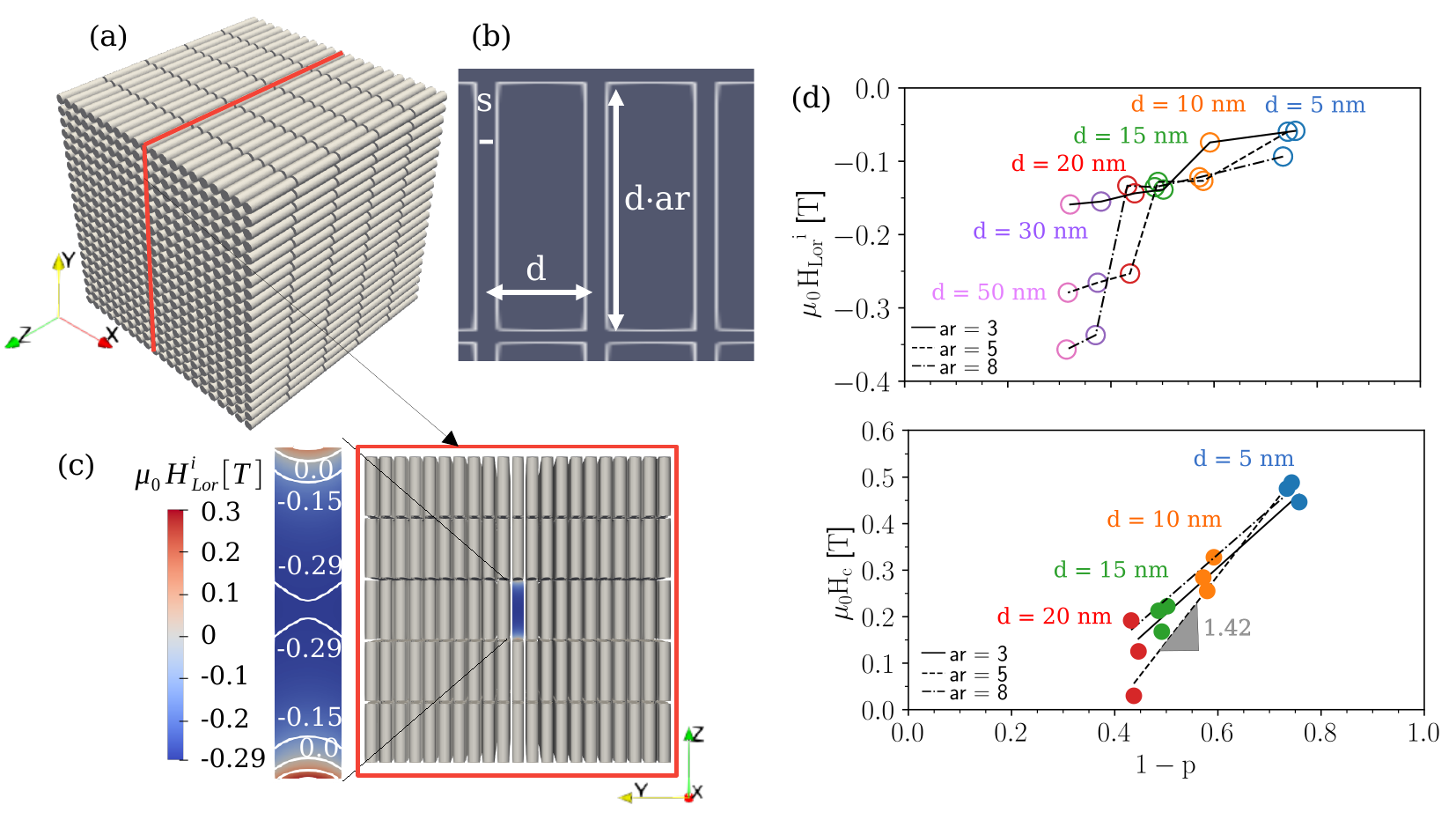}
 \caption{Schematic of the arrangement of 2204 rods, with their longitudinal axes aligned along the z-axis (a). The geometric parameters of the rods (b). A central cross-section in the YZ plane through the arrangement of rods and a map of the vertical component of the computed stray field ($B_z$). $B_z$ contours are shown on the map, with values indicated by the color bar (c). The average of $B_z$ over all grid points as a function of p (d). Coercivity estimated on the basis of $B_z$ as a function of p. The slope of the linear fit of coercivity versus packing fraction for an aspect ratio of 5 is shown in the plot (e). In (d) and (e) diameters are distinguished by color and aspect ratios by line style.}
\label{fig11}
\end{figure}

We computed the stray field generated on a centred rod by surrounding neighbours by modelling each rod as an ideal, azimuthally symmetric solenoid  \cite{Derby_AJP2010}. The number of surrounding neighbours was chosen as a threshold beyond which further increase will not change the stray field generated on the centred rod. Figs. \ref{fig11} (a) and (b) show the arrangement of 2204 rods and their geometric parameters, respectively.

Fig. \ref{fig11} (c) shows a central cross-section through the arrangement of rods and a map of the vertical component of the computed stray field ($B_z$) which was used to calculate the coercive field of the ensemble of rods. We exemplified with \SI{20}{nm} thick rods and \SI{5}{nm} interrod spacing, similar to the geometric parameters observed in the reference sample. We adjusted the coercive field of the multi-rod system by the computed $B_z$ averaged over all grid points. The average of $B_z$ over all grid points is supported by similar value of the coercivity obtained with this method as in the case of finite element micromagnetic simulations of a single-rod with a precomputed field added to the applied field. The role of the precomputed field is to account for the magnetostatic interactions. The average of $B_z$ over all grid points is shown in Fig. \ref{fig11}(d) for different packing fractions obtained by varying the dimensions of the rods. The magnitude of the stray field generated by neighbouring rods increases with the aspect ratio and diameter, consistent with the corresponding reduction in coercivity. For diameters greater than \SI{20}{nm}, the increase in $B_z$ becomes more pronounced with increasing aspect ratio than with increasing rod diameter. Fig. \ref{fig11} (e) shows the dependence on packing fraction of the coercivity estimated on the basis of $B_z$ for various rod diameters and aspect ratios. For each aspect ratio considered, this dependence was analysed by an ordinary least squares linear regression. The fit yielded a slope of 1.42 and a coefficient of determination of about 0.98 for an aspect ratio of 5. With the linear dependence of coercivity on packing fraction established, we scaled the coercivity calculated for the multi-rod system by $(1-p) $ in subsequent calculations. This analytical calculation confirms that the average stray field acting on a central rod increases approximately linearly with the packing fraction and, therefore, reduces the coercive field in a way consistent with the dependence. While the simplified theoretical model for non-interacting single-domain particles suggests that coercivity should be directly proportional to $(1-p)$ and thus reach zero exactly at a packing fraction of $p=1$ , this ideal behaviour is not reflected in real-world materials. Historical experimental data \cite{Luborsky_JAppPhys1957, Luborsky_JAP1961} demonstrates that coercivity declines much faster than the theoretical equation predicts, typically extrapolating to $H_c \sim 0 $ at roughly $p \sim 0.7$. Our micromagnetic simulations show a similar trend. Extrapolating to lines for the coercive field as function of (1-p) in Figure \ref{fig11} (d) shows that coercivity reaches zero at about a packing fraction of 0.7.

\end{document}